\documentclass[amssymb,aps,prl,a4paper,showpacs]{revtex4} 
 
\usepackage{epsfig,amsfonts,amssymb,amsmath} 
\begin{document} 
\topmargin=0.2cm 
\title{Interacting Geodesics: Binary Systems around a Black Hole.}

 \author{Eduardo Gu\'eron}\email[e-mail: ]{gueron@ime.unicamp.br}

 \author{Patricio S. Letelier}\email[e-mail: ]{letelier@ime.unicamp.br} 
\affiliation{ 
Departamento de Matem\'atica Aplicada, Instituto de Matem\'atica, 
Estat\'{\i}stica e Computa\c{c}\~ao Cient\'{\i}fica, Universidade 
Estadual de 
Campinas, 13083-970, Campinas, SP, Brazil 
} 


\begin{abstract}

We present a novel method to study interacting orbits 
in a fixed mean gravitational field associated with a solution 
of the Einstein field  equations. The idea is to consider the Newton gravity
among the orbiting particles in a geometry given by the main source. 
We apply the technique in the of study  two and three self-gravitating
particles moving around a black hole, i.e., in a Schwarzschild
geometry. We also compare with  the equivalent Newtonian
problem and noted differences in the structural stability, e.g.,
binary systems were found only in the general 
relativistic approach.

\end{abstract} 
\maketitle  

General relativistic effects are usually neglected in astrophysical 
systems in which many gravitational sources are considered.  In most 
 of  the known examples these  effects are insignificant. Also 
 the n-body problem  is intractable in  Einstein theory without large 
approximations. 
Nevertheless the difference between the Newtonian  and the General 
Relativistic approaches  becomes important close to compact objects and 
around very massive structures. When non interacting particles are 
been considered, the problem can be solved computing 
time-like geodesics. When more than one source is considered (in the 
non-stationary case) numerical techniques might be used to 
 solve the Einstein field equations. However finding numerical 
solution to the simple two body problem is already a non trivial 
problem, see for instance \cite{marronetti}. 
In some cases, analytical and numerical solutions to the n-body 
problem in GR  are found 
as sequences  of post-Newtonian approximations 
\cite{spyrou}. Following the post-Newtonian idea, there are some
predictions of relativistic effects using a relativistic celestial
mechanic \cite{mashhoon}. Other  approximation techniques are also 
available,  the simplest  consists 
in  solving a typical Newtonian problem 
using pseudo-potentials that simulates the GR aspects relevant in the 
studied situation \cite{pw, sk}.

We suggest in this article a novel technique that makes possible the 
study of small  massive objects orbiting around  a fixed 
structure with a large mass. The idea 
is to consider the Newtonian gravity among particles evolving in a 
geometry associated with  a solution to the Einstein field 
equations.

We construct  within the  Newtonian  gravity a quadri-force. Then this 
 force is used to modify the geodesic equation to allow 
a ``Newtonian'' interaction among them. In this scenario the damping 
 due to the emission of gravitational radiation is not considered.

In the context of Schwarzschild geometry, we study planar 
interacting orbits  and compare the results with the equivalent situation in 
the usual Newtonian theory. We see a great difference in the 
 structure stability 
of the studied systems.  The most relevant   result is that in our 
  approach the generation of binary systems is  favoured if compared with 
the equivalent pure Newtonian situation. In n-body simulations the production 
 of binary systems is a key issue in the formation of 
 astrophysical  structures as galaxies and star clusters \cite{spur}.


In our proposition the interacting particles do not change the mean 
gravitational field. In the general relativistic context this field is 
represented by a solution to the Einstein 
equations. Non-interacting particles  move following geodesics, i.e., their 
world lines obey  the geodesic equations, 
\begin{equation} 
\dot x^\lambda \nabla_\lambda \dot x^\mu  \equiv 
\ddot x^\mu+\Gamma_{\alpha\beta}^\mu\dot 
x^\alpha \dot x^\beta=0, \label{geod} 
\end{equation} 
where $\Gamma_{\alpha\beta}^\mu=g^{\mu\gamma}(g_{\alpha\gamma,\beta}+g_{\beta\gamma, 
\alpha}- g_{\alpha\beta,\gamma}),$ 
and $g_{\alpha\beta}$ is the metric tensor for the known 
solution. The  dots represent  ordinary derivation with respect 
 to the proper time 
$\tau$ and commas denote the 
 usual partial derivation with respect to the coordinate.

The next step is to 
use Newton gravity to 
compute the interaction among  orbiting particles. The Newtonian 
gravitational potential generated by a set of $N$ massive particles is     
\begin{equation} 
\Phi=- \sum_{a=1}^N\frac{G m_a}{|{\bf r}-{\bf r}_a|}.\label{pot} 
\end{equation} 
In  the rest of the article we shall use geometrical 
 units: the Newton constant $G=1$, the light speed $c=1$,  we also assume that the black hole 
mass is unit ($M=1$).

Therefore the force ${\bf f}$ that acts on the particle $b$ due to the other 
orbiting particles is 
\begin{equation} 
{\bf f}_{(b)}=-m_b {\boldsymbol\nabla_b} \Phi({\bf r}_a\ne{\bf r}_b ).\label{fnewt} 
\end{equation} 
Now we  look for a 
quadri-force that represents in some way the Newtonian interaction 
among the particles in a fixed pseudo-Riemannian geometry. 
The spatial components of the quadri-force $F^\mu$ is 
  obtained from ${\bf f}$ given in (\ref{fnewt}). From  ${\bf 
  f}=d{\bf p}/dt$ and $F^i=dp^i/d\tau$ we get $F^i=u_0 f^i$, 
where $u^\mu=dx^\mu/d\tau$. 
We want  a   quadri-force $F^\mu$   orthogonal to the 
quadri-velocity $u^\mu$. Hence, we use  $u_\mu F^\mu=u_0F^0+u_0u_i
f^i=0$ to obtain $F^0=-u_i f^i$. We can write $F^\mu$ in a more
adequate form as  
\begin{eqnarray} 
F^i=m u^0g^{ij}\frac{\partial \Phi}{\partial x^j},& 
F^0=-m g^{00}u^i\frac{\partial \Phi}{\partial x^i}.\label{rgforce} 
\end{eqnarray} 
The signature of $g_{\mu\nu}$ is taken as $-2$.

Finally, we introduce $F^\mu$ in the right side of the geodesic 
equations representing an external force to define the motion 
equations for a   particle of mass $m_a$, 
\begin{equation} 
m_a \dot x^\lambda_{(a)}\nabla_\lambda \dot
x^\mu_{(a)}=F_{(a)}^\mu. \label{aig}  
\end{equation}  
 From now on we consider that all the  particles 
moving around the center of attraction have the same  mass $m$.

First we consider  the limits of  (\ref{aig}) for 
large and small masses. When the mass of the interacting particles $m$
is much smaller than the fixed source mass $M$, we have nothing but the
geodesic equations for a set of  
test-particles. For $m\gg M$ we get the special relativistic analogous to the 
Newton second law.   And for small speeds, i.e., $u_0\rightarrow 1$ and 
$u_i\rightarrow 0$, the Newtonian motion equations for a system of $N$ 
gravitating particles are recovered.


The method is ``self-inconsistent'' since some of the mathematical
principles of 
a general relativistic framework are not satisfied. For instance, we are using 
an instantaneous interaction among the particles. The distance
between two particles calculated in the Black Hole framework was
used to compute the force of attraction without correction
related to the different speed of the particles. The force is not
covariant. 

We must have these imperfections in mind to determine the limitation of the
method and to check numerical results. Since the geometry will be kept
fixed the mass of
the orbiting particles must be much smaller than the main source. We
have to systematically control 
variations of the total angular momentum and energy of the orbiting
particles, they should be constant.


We use  the above described approach 
 to study  particles moving around a spherical symmetric source, i.e, 
we  numerically solved (\ref{aig}) when the field $g_{\mu\nu }$ is 
 the Schwarzschild metric.  
The initial conditions are chosen  to have for each particle  a  timelike 
worldline and  that the  spatial projection of these worldlines 
 represent a  bounded system of particles moving on the 
 plane $\vartheta=\pi/2$. 
 
In the first examples, three particles moving close to the black hole
 are studied. For this purpose we take : $L_1=6.752, E_1=0.988; 
 L_2=6.937, E_2=0.989$ and $L_3=5.895, E_3=0.984$. [In the
 Schwarzchild metric: $L=r^2 {\dot\varphi}\sin^2\vartheta$ and $E={
 \dot t} /(1-2M/r)$]. Their initial distances from the center of
 attraction are, respectivily, $r_1=42.5, r_2=45.0$ and $r_3=31.5$ in
 geometric units.

\begin{figure}[p] 
\epsfig{width=2in,height=1.5in, file=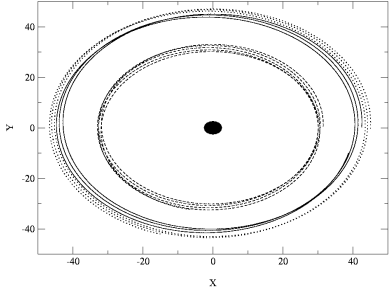}  
\vspace{0.2cm}
\caption{Geodesic motion of three particles around the black hole. The
  initial angular momentum and energy of each particle are,
  respectively,  $L_1=6.752, E_1=0.988; 
 L_2=6.937, E_2=0.989$ and $L_3=5.895, E_3=0.984$.}   
\label{rg0_0} 
\end{figure}

\begin{figure}[p] 
\epsfig{width=2in,height=1.5in, file=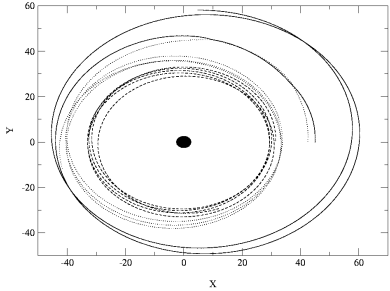}  
\vspace{0.2cm} 
\caption{Now we use the same values of angular momentum and energy of
 the first figure but now with interacting particles. They have the
 same mass, $m=5\times 10^{-4}$. We obtain three particles in a
 bounded motion around the black hole.}    
\label{rg5_4} 
\end{figure}

\begin{figure}[p] 
\epsfig{width=2in,height=1.5in, file=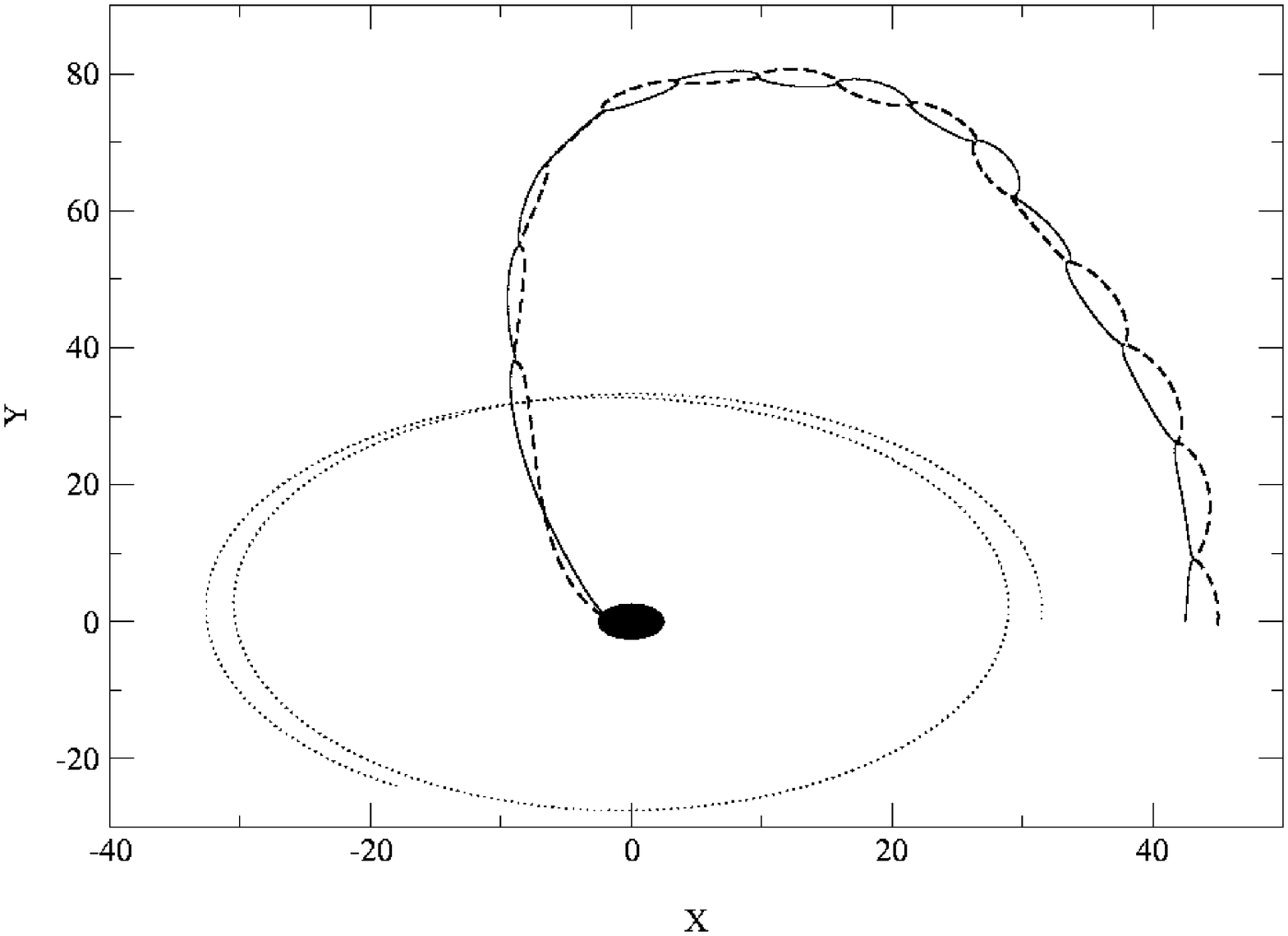}
\vspace{0.2cm}   
\caption{Using the same values of angular momentum and energy of the
  preceding figure and $m=3\times 10^{-3}$ we have now that two of the
  particles form a binary system that falls into the black hole while
  the other stabilyzes in a bounded motion around the center.}   
\label{rg3_3} 
\end{figure}

 When we compute the orbits for three  particles interacting only with a 
 center of attraction   ($F_{(1)}^\mu= F_{(2)}^\mu=F_{(3)}^\mu=0 $) this is
 equivalent to chose  $m=0$, as above mentioned we take the black hole mass 
as one  ($M=1$). For the  
 already presented  values of $L$ and $E$ we have  bounded  geodesic motion
 around the black hole Fig.\ref{rg0_0}. In  
Fig.\ref{rg5_4} we present three particles with equal masses 
 $m=5\times 10^{-4}$. The trajectories are quite different from the
 geodesics  but the particles 
still have bounded motion. In Fig.\ref{rg3_3} we present the motion for 
 a larger value of the masses, $m=3\times 10^{-3}$. In this case, one   
particle escapes (the dashed orbit) while the other couple form a
binary system that falls into the black hole.

\begin{figure}[p] 
\epsfig{width=2in,height=1.5in, file=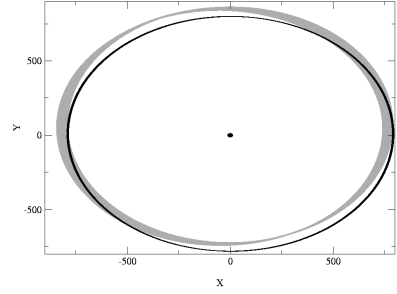} 
\vspace{0.2cm}  
\caption{Geodesic motion around a black hole of two  particles with $L_1=28.10,
  E_1=0.99937$ and  $L_2=28.16, E_2=0.99937$.} 
\label{rg0d0} 
\end{figure}

\begin{figure}[p] 
\epsfig{width=2in,height=1.5in, file=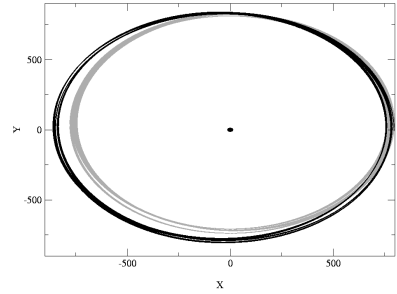} 
\vspace{0.2cm}  
\caption{We maintain the values of angular momentum and energy of the
  preceding figure with two interacting particles with $m=10^{-5}$. They
  have a bounded motion close to the geodesics.} 
\label{rg1d5} 
\end{figure}

\begin{figure}[p] 
\epsfig{width=2in,height=1.5in, file=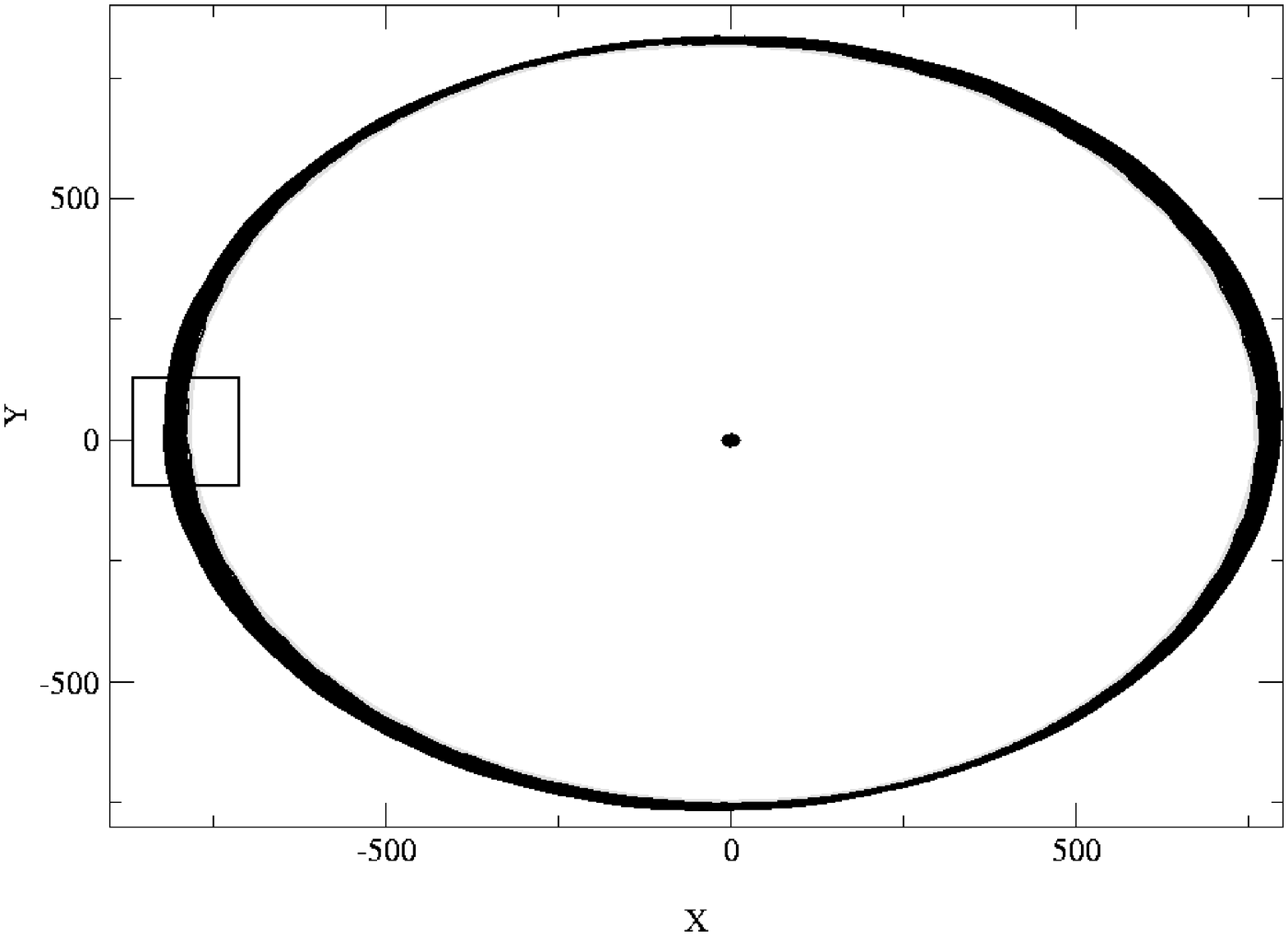}  \hspace{0.5cm}  
\vspace{0.2cm}
\epsfig{width=1.08in,height=0.81in, file=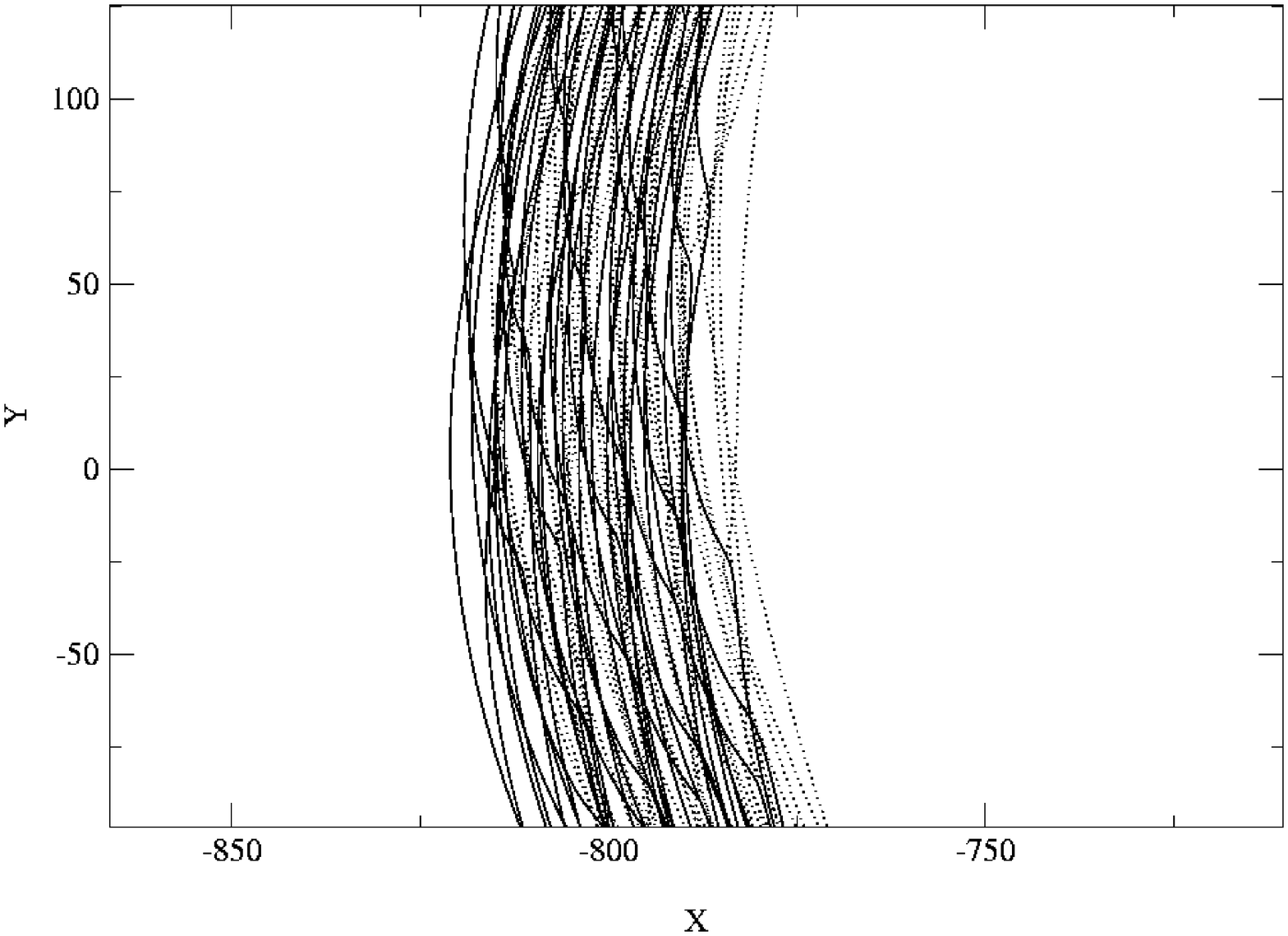} 
\caption{Using the same $L$ and $E$ of the last figure but with
  $m=10^{-4}$, we have that 
  the motion of the particles is still bounded but now they are more
  irregular and we cannot easily distinguish them.} 
\label{rg1d4}
\vspace{1cm} 
\end{figure}

\begin{figure}[h] 
\epsfig{width=2in,height=1.5in, file=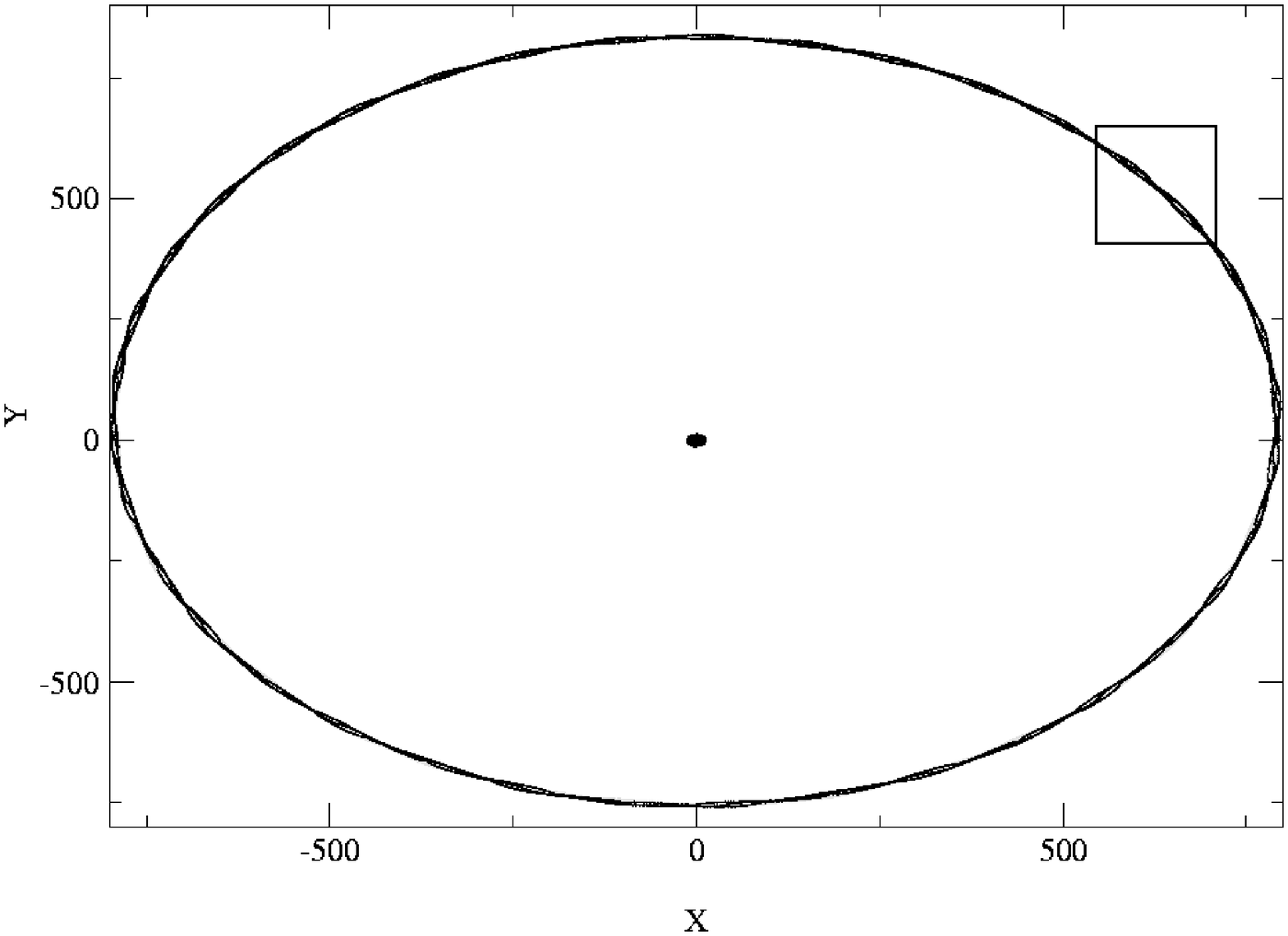} \hspace{0.5cm}  
\vspace{0.3cm}
\epsfig{width=1.08in,height=0.81in, file=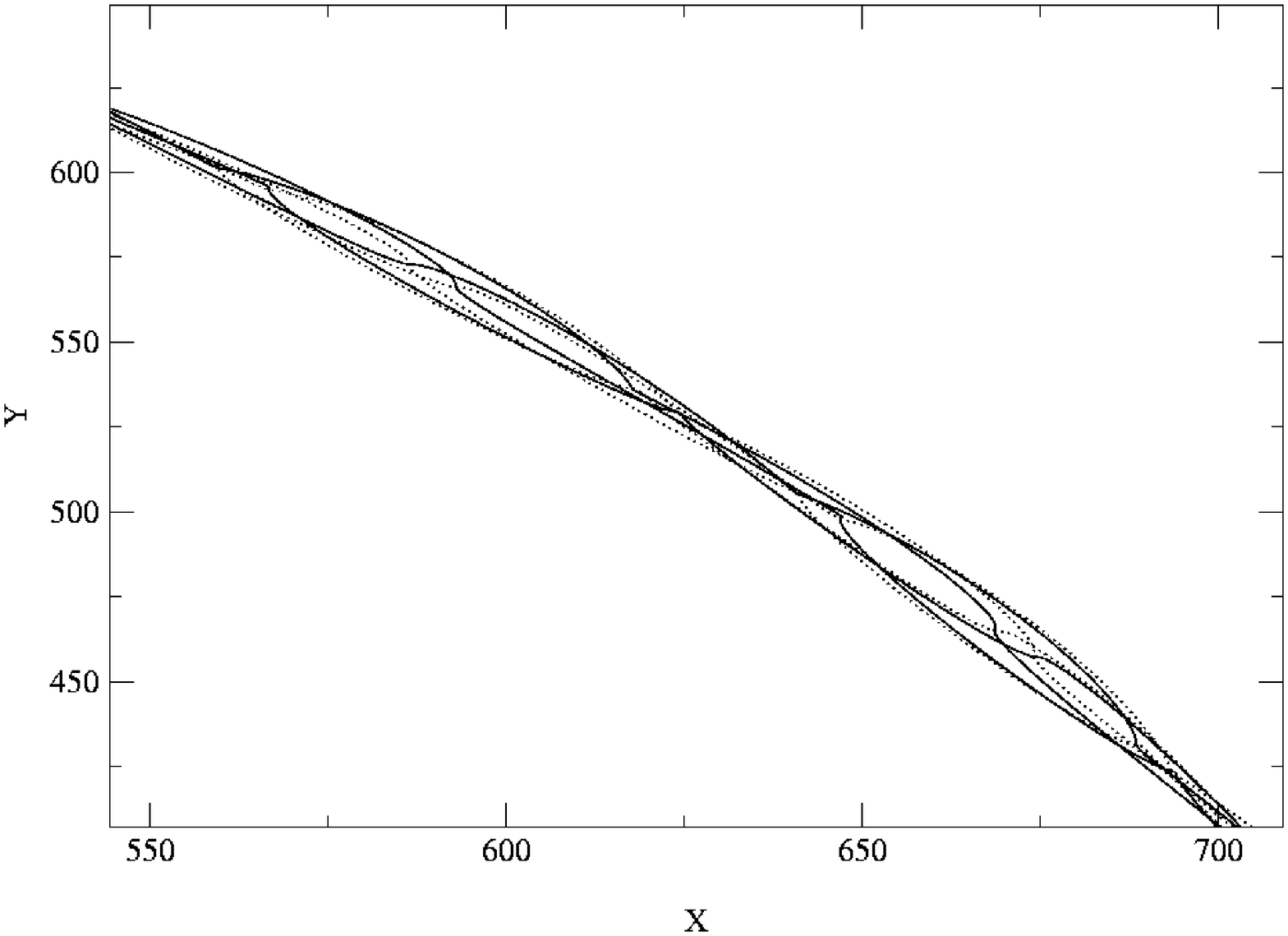}  
\caption{Increasing the mass to $m=5\times 10^{-4}$ we obtain a binary
  system moving around the center of attraction.} 
\label{rg5d4} 
\end{figure}

Now we  consider only two particles farther from the black hole than
  the preceding ones. For this purpose, we consider $L_1=28.10,
  E_1=0.99937$ and  $L_2=28.16, E_2=0.99937$. As in the previous
  examples, the non interacting  orbiting particles follow bounded
  geodesic motions, Fig.\ref{rg0d0}. In Fig.\ref{rg1d5} we present the
  trajectories  
 for the pair of orbiting particles with $m=10^{-5}$. In this case,
  we can clearly  dintinguish the motion of 
 each particles. In the second case is considered $m=10^{-4}$, now the
  particles move  in the same region of the space in such
 way that their orbits fill a thick ring, Fig.\ref{rg1d4}. Finally, in
  Fig.\ref{rg5d4},  the particles mass are  $m=5\times
 10^{-4}$  and we have the  formation of  a stable binary 
system that remains in a confined motion around the black hole.

It  is instructive to compare our results with the equivalent 
pure Newtonian  three-body problem. 
The  gravitational potential field of the system in this case is 
\begin{equation} 
\Phi_{Newt}({\bf r})=-\frac{GM}{r}-\sum_{a=1}^{2}\frac{Gm_a}{|{\bf 
r}-{\bf r}_a|}.\label{phinewt} 
\end{equation} 
With $m_1=m_2=m$ the motion equations of the particles are, 
\begin{equation} 
\frac{d^2{\bf x}_a}{dt^2}=-\nabla_a\Phi_{Newt}.\label{eqmot} 
\end{equation} 
These  equations 
 are numerically solved 
 by using parameters that confine the   
particles in a motion around the central mass. 
The Newtonian  energy per unit of mass, $E_{Newt}$, is related to the
 relativistic  
specific energy by   $E=\sqrt{1-2E_{Newt}}$. 
The specific angular momentum $L$ and  energy $E$ used in the 
simulations are  $L_1=6.55, E_1=0.988; L_2=6.75, E_2=0.989,$ and
 $L_3=5.68, E_3=0.984$. These  
 values are equivalent to the ones used in the first general relativistic   
 simulations (Figs.\ref{rg0_0}, \ref{rg5_4} and \ref{rg3_3}).

\begin{figure}[h] 
\epsfig{width=2in,height=1.5in, file=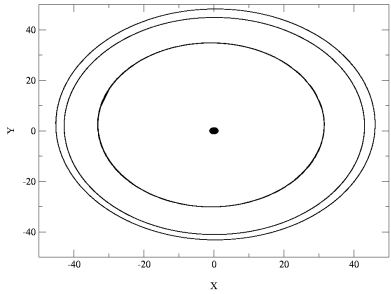} 
\vspace{0.3cm}  
\caption{Newtonian analogous to Fig.\ref{rg0_0}. Three test-particles with
  $L_1=6.55, E_1=0.988; L_2=6.75, E_2=0.989.$ and $L_3=5.68,
  E_3=0.984$. } 
\label{nt0_0} 
\end{figure}

\begin{figure}[p] 
\epsfig{width=2in,height=1.5in, file=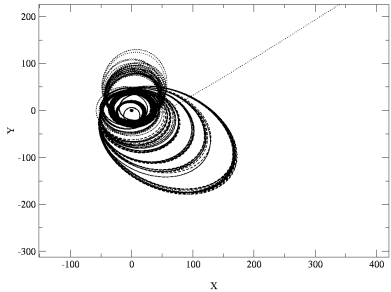}
 \vspace{0.3cm}  
\caption{With the same values of $L$ and $E$ of the preceding figure,
 and $m=5\times 10^{-4}$ we get the Newtonian analogous to
 Fig.\ref{rg5_4}.
 We can see that one of the particles escapes. } 
\label{nt5_4} 
\end{figure}

\begin{figure}[p] 
\epsfig{width=2in,height=1.5in, file=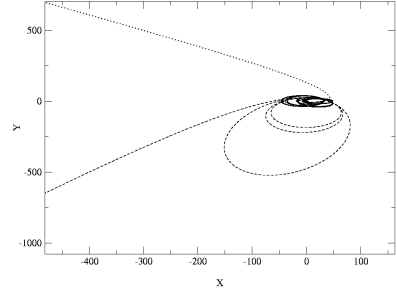}
\vspace{0.3cm} \vspace{2cm}
\caption{Increasing the mass, $m=3\times 10^{-3}$, we have that two of
  the particles quickly escapes from the system while the remaining
  one tends to stabilizes in a elliptical motion.} 
\label{nt3_3} 
\end{figure}

We start with the simulation of three test-particles   
(no interaction among them) moving around an attraction center 
 in elliptical orbits, Fig.\ref{nt0_0}. We present in Fig.\ref{nt5_4}
 interacting 
 particles with  $m=5\times 10^{-4}$. We see that one of them quickly
 unbound and the remaining couple stay in a bounded motion
 for some time. With a stronger interaction, $m=3\times 10^{-3}$
 presented in Fig.\ref{nt3_3}, two
 particles quickly 
move far from the center of attaction while the remaining one stabilizes
 in a bounded elliptical motion closer to the black hole.

\begin{figure}[p] 
\epsfig{width=2in,height=1.5in, file=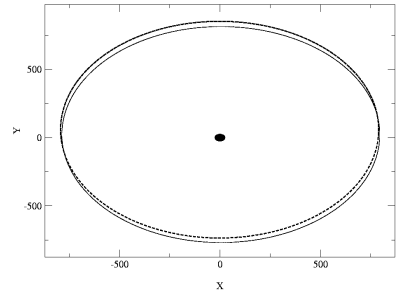} 
\vspace{0.3cm}
\caption{The  Newtonian analogous 
  to the system presented in Fig.\ref{rg0d0} is reproduced here. Two
  test-particles with  $L_1=28.10, E_1=0.99937$ and
$L_2=28.12, E_2=0.99937$. The
  Newtonian analogous 
  to the system presented in Fig.\ref{rg0d0} is reproduced.} 
\label{newt0_0} 
\end{figure}

\begin{figure}[p] 
\epsfig{width=2in,height=1.5in, file=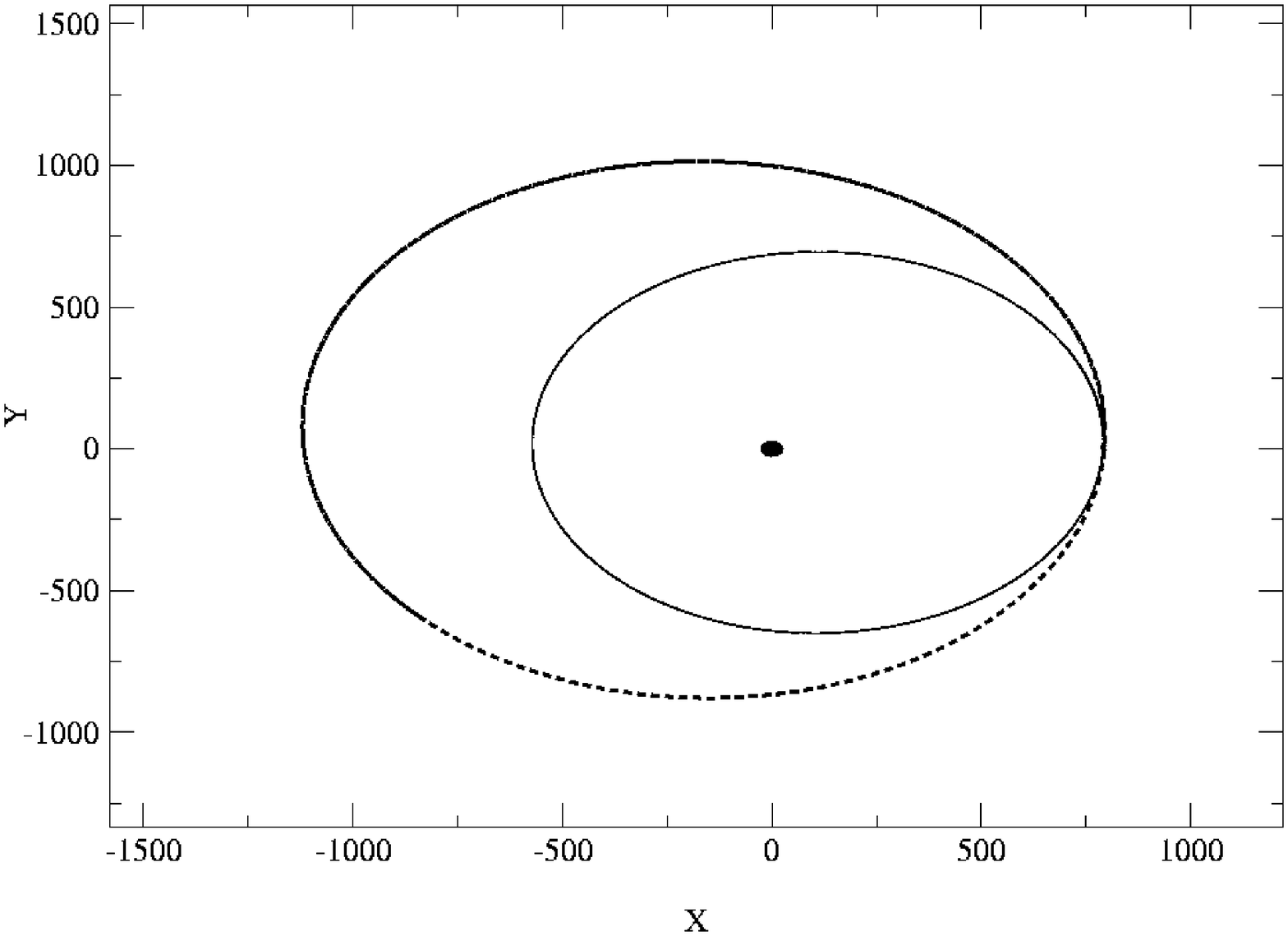}
\vspace{0.2cm} 
\caption{We use the same values of angular momentum and energy of the
  preceding figure with interacting particles with $m=10^{-4}$. The
  Newtonian analogous 
  to the system presented in Fig.\ref{rg1d4} is reproduced. We obtain,
  as a final 
  configuration, two quasi-elliptical motion.} 
\label{newt1_4} 
\end{figure}

\begin{figure}[t] 
\epsfig{width=2in,height=1.5in, file=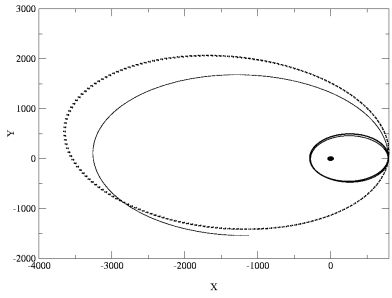} 
\vspace{0.2cm}
\caption{Using the same values of $L$ and $E$ as in the previous figure
  and increasing to $m=10^{-3}$ we obtain that the motion of the
  particles are linked in such way that they alternate between a small
  and a large precessing ellipse.}  
\label{newt1_3} 
\end{figure}

Now we present results of two particles moving farther from the
center of attraction in a situation similar to the analogous
relativistic presented in Fig.\ref{rg0d0}, \ref{rg1d5}, \ref{rg1d4} and
\ref{rg5d4}. For this aim, we use $L_1=28.10, E_1=0.99937$ and
$L_2=28.12, E_2=0.99937$. The motion of test particles is presented in
Fig. \ref{newt0_0}. In Fig.\ref{newt1_4} we have $m=10^{-4}$ the
particles stabilize 
in quasi-elliptical orbits. One closer and other more distant to the
center of mass if we compare with the non-interacting situation. For a
stronger interaction, $m=3\times 10^{-3}$, we have a interesting
situation as shown in Fig.\ref{newt1_3}. The particles alternate
quasi-elliptical motions closer an farther to the center of
attraction, i.e., while one of them are orbtitating in a large ellipse
the other is in a small one. We make some simulations considering that
the central mass is not fixed and has the unit mass ($M=1$). No
significative difference was noticed.

Before analysing the results, it is important to explain some
numerical cautions necessary in these general relativistic
simulations. In the first examples (Fig.\ref{rg5_4} and Fig.\ref{rg3_3})
the particles were very close to the   black hole (remember that if
the BH have $10M\odot$ the distance of  
 $50$ in geometric units will be about $50km$)  therefore the system is more
unstable, see for instance \cite{gueron,letelier}, and the error due to the
inexact approach is larger since the general relativistic effects are
very important. We use, to control such problems, the numerical values
of the total angular momentum and energy of the orbiting particles (that should
be constant) and their physical time - if they were very
different we would have problems in the syncronization of the
Newtonian interaction. The numerical variation of the total energy was
always smaller than $1\%$, the total angular momentum varied at most $10\%$   
and the time less than $5\%$. We consider that it is the worst
situation that we still can apply this technique, in the other
examples with the general relativistic approach, Fig.\ref{rg1d5} and
Fig. \ref{rg1d4} the deviations of the total angular momentum, total
energy and time were very small - always less than $1\%$.


By comparing the orbits obtained in our GR framework 
 with the equivalent situation in Newtonian gravity 
 we  conclude that the strutuctual stability of the same system
 changes substancially with different approaches. The most interesting
 difference is the formation of binary systems noticed in the Fig.\ref{rg3_3}
 (a binary system that is captured by the black hole) and in the
 Fig.\ref{rg5d4} (the binary bounded motion stabilizes around the
 black hole). An analogous situation is very hard to be obtained in the
 Newtonian approach so close to the center of attraction. When the
 orbiting particles are very far from the main gravitational source,
 the Newtonian and the relativistic approach lead to similar
 results. In these cases, the gravitational force
 exerced by the central mass varies much more slowly than the
 interaction between the two orbiting particles.

\acknowledgments 
{We thank to FAPESP for the financial support. E.G.  also thanks 
Prof. Samuel Oliveira for discussions about numerical problems and P.S.L. 
also thanks CNPq.}

\twocolumngrid

\vspace{2cm}

\end{document}